\documentclass[letter,twocolumn]{jpsj2} 
\title{Pressure-temperature Phase Diagram of Polycrystalline UCoGe Studied by Resistivity Measurement}

\author{Elena \textsc{Hassinger}\thanks{E-mail: elena.hassinger@cea.fr}, Dai \textsc{Aoki}\thanks{E-mail: dai.aoki@cea.fr}, Georg \textsc{Knebel}, and Jacques \textsc{Flouquet}}

\inst{Institute for Nanoscience and Cryogenics, SPSMS\\ CEA Grenoble, 17 rue des Martyrs, 38054 Grenoble Cedex 9, France}

\abst{
Recently, coexistence of ferromagnetism ($T_{\rm Curie}=2.8$~K) and superconductivity ($T_{\rm sc}=0.8$~K) has been reported in UCoGe, a compound close to a ferromagnetic instability at ambient pressure $P$.
Here we present resistivity measurements under pressure on a UCoGe polycrystal. The phase diagram obtained from resistivity measurements on a polycrystalline sample is found to be qualitatively different to those of all other ferromagnetic superconductors. By applying high pressure, ferromagnetism is suppressed at a rate of 1.4~K/GPa. No indication of ferromagnetic order has been observed above $P\approx1$~GPa. 
The resistive superconducting transition is, however, quite stable in temperature and persists up to the highest measured pressure of about 2.4~GPa. Superconductivity would therefore appear also in the paramagnetic phase. However, the appearance of superconductivity seems to change at a characteristic pressure $P^\star \sim 0.8$~GPa. Close to a ferromagnetic instability, the homogeneity of the sample can influence strongly the electronic and magnetic properties and therefore bulk phase transitions may differ from the determination by resistivity measurements. 
}

\kword{UCoGe, ferromagnetic superconductor, high pressure}

\begin{document}
\maketitle
Since superconductivity under pressure ($P$) in the ferromagnet UGe$_2$ has been discovered \cite{Saxena2000}, the coexistence of ferromagnetism and superconductivity has attracted much attention because in standard BCS theory spin singlet pairing is unlikely in a ferromagnet. It had been proposed \cite{Fay1980} that superconductivity could occur in a weak itinerant ferromagnet at the border of the magnetic state when the attraction of electrons forming a Cooper pair is mediated via magnetic fluctuations. In this case, the Cooper pair electrons would form a triplet ground state. Up to now, coexistence of bulk superconductivity and ferromagnetism has been reported in UGe$_2$ \cite{Saxena2000}, URhGe \cite{Aoki2001}, and most recently in UCoGe \cite{Huy2007}. 
In UGe$_2$ superconductivity appears only under pressure in the ferromagnetic state in a pressure window from 1~GPa up to the pressure $P_{\rm c}\sim1.5$~GPa, where the ferromagnetism collapses by a first order transition. At $P_{\rm x}=1.2$~GPa a transition between two different ferromagnetic states (FM1 and FM2) occurs with a change of the ordered moment at $T=0$ from $m_0\sim 1.4$~$\mu_{\rm B}$ to $m_0\sim 1$~$\mu_{\rm B}$ \cite{Tateiwa2001, Pfleiderer2002}. Even close to the critical pressure, the magnetic moment is still large. The temperature of superconductivity $T_{\rm sc}$ reaches its maximum (700~mK) at this pressure $P_{\rm x}$ \cite{Tateiwa2001a}.  This strongly suggests that superconductivity is rather enhanced by the fluctuations at the transition between FM1 and FM2 than at the quantum critical point. 

In URhGe superconductivity occurs already at ambient pressure in the ferromagnetic phase ($T_{\rm Curie}=9.5$~K, $T_{\rm sc}=250$~mK, $m_0=0.4$~$\mu_{\rm B}$) \cite{Aoki2001}.  Applying pressure tunes the system away from the quantum critical point. An almost linear increase of $T_{\rm Curie}$ with a slope of $dT_{\rm Curie}/dP= 0.65$~K/GPa up to 12~GPa \cite{Hardy2005} is observed under pressure, but the superconducting transition temperature decreases. Additionally, a spectacular reentrance of superconductivity appears close to a reorientation of the magnetization under high magnetic field \cite{Levy2005}.

These two cases of ferromagnetic superconductors give no clear view on the link between superconductivity and a first order transition of a weak ferromagnetism at $P_{\rm c}$. 
The recently discovered  material UCoGe, which is also ferromagnetic and superconducting at ambient pressure ($T_{\rm Curie}=2.8$~K, $T_{\rm sc}=0.8$~K) but has a lower sublattice magnetization of $m_0=0.07$~$\mu_{\rm B}$ than the previous systems \cite{Huy2007, Huy2008}, seems to correspond to the required condition. Furthermore, from thermal expansion and specific heat measurements on UCoGe it was predicted by applying the Ehrenfest relation that $dT_{\rm Curie}/dP=-2.5$~K/GPa and $dT_{\rm sc}/dP=0.48$~K/GPa \cite{Huy2007}. Assuming a linear $P$ variation $T_{\rm Curie}(P)$ a critical pressure of $P_{\rm c} \approx1.2$~GPa is expected for the case of a second order quantum phase transition at $T=0$. 
Our main aim was to determine the pressure dependence of $T_{\rm sc}$ and $T_{\rm Curie}$ in the proximity to this pressure.

\begin{figure}
\begin{center}
 \includegraphics[width=76mm]{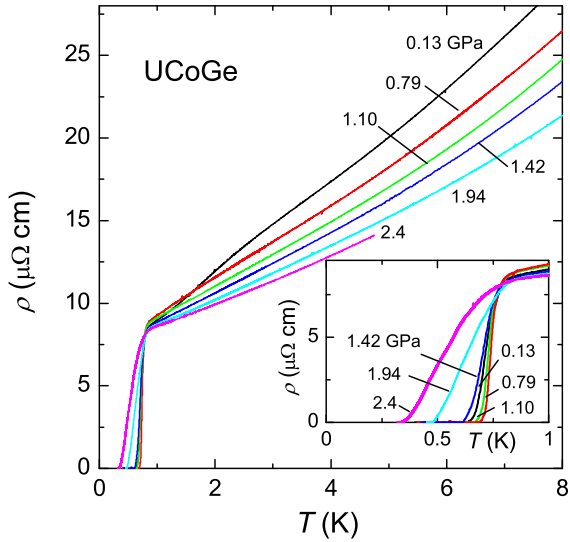}
\end{center}
 \caption{(Color online) Resistivity $(T, P)$ of UCoGe for different pressures between 0.13~GPa and 2.4~GPa. The inset shows a zoom into the low temperature region with the superconducting transition, which is seen for all measured pressures.}
 \label{data}
\end{figure}

Extensive pressure studies via resistivity measurements are reported. A polycrystalline sample was prepared by using a radio frequency furnace. The starting materials U (purity 99.9 \%, 3N), Co (4N) and Ge (6N) were melted in a water cooled copper crucible under ultra high vacuum (UHV) and purified argon atmosphere. The sample was annealed for 5 days at 900 $^\circ$C under UHV. X-ray diffraction confirmed the TiNiSi-type orthorhombic crystal structure. The small sample for pressure measurements has been cut by spark cutter to a size of about $0.5\times 1 \times 1$~mm$^3$. The residual resistivity ratio ($RRR=\rho$(300~K)/$\rho$(1K)) is approximately 28, indicating the high quality. At ambient pressure the polycrystalline sample has ferromagnetic and superconducting signatures quite similar to those found in Ref.~\citen{Huy2007} but slightly broader than in single crystals \cite{Huy2008}.

Resistivity was measured by standard four point method with a lock-in amplifier at a measurement frequency of $f=17$~Hz with an electrical current of $\sim$ 25~$\mu$A. The signal was amplified by a factor of 100 with an transformer at ambient temperature and by a factor of 1000 with a low noise pre-amplifier. Pressure was applied by a hybrid piston cylinder cell with a NiCrAl inner cylinder with a inner diameter of 4~mm and CuBe outer cylinder. The feedthrough and piston are of CuBe and tungsten carbide, respectively. Daphne oil 7373 has been used as a pressure transmitting medium and the pressure was determined via the measurement of the ac-magnetic susceptibility of the superconducting transition of a piece of lead inside the pressure chamber. At constant pressure temperature sweeps at zero field have been effected in a $^3$He cryostat down to 0.4~K and for the three highest pressures in a dilution refrigerator down to 0.1~K. 

\begin{figure}
\begin{center}
 \includegraphics[width=76mm]{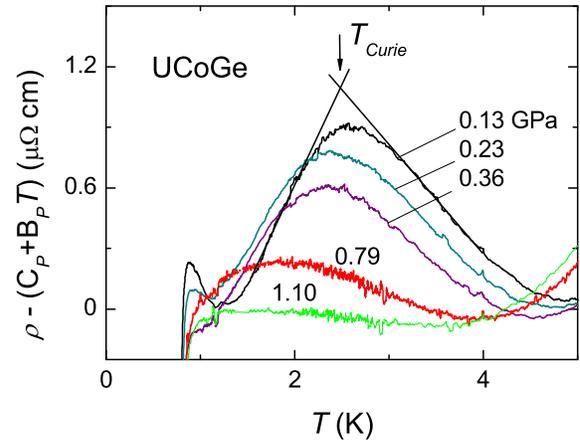}
\end{center}
 \caption{(Color online) Determination of the Curie temperature $T_{\rm Curie}$ in UCoGe: A straight line (resistivity linear in $T$) has been subtracted from the resistivity data for each pressure to see clearly the transition temperature which has been determined by the crossing point of the two tangents; here presented for $P=0.13$~GPa.}
 \label{minusgerade}
\end{figure}

Fig. \ref{data} shows the resistivity data as a function of temperature for different pressures. The absolute value of the resistivity is normalized to $\rho=250$~$\mu\Omega\,$cm at room temperature, which was estimated from the resistivity of URhGe. The appearance of micro cracks in the polycrystalline sample impedes the determination of the resistivity by the geometrical factor. At the lowest pressure, the anomaly at $T_{\rm Curie}(\rho)$ is visible as a very small broad anomaly, which becomes even less pronounced for higher pressures. 
Therefore the ferromagnetic transition temperature is difficult to define. The maximum in the temperature derivative of the resistivity curves does not provide a reasonable criterion to locate $T_{\rm Curie}(\rho)$. The preferable method we found was to subtract a straight line ($C_{\rm P}+B_{\rm P}T$) from each resistivity curve. The obtained curves up to 1.1~GPa are shown in Fig. \ref{minusgerade}. In that way, the transition is clearly visible and can be defined by a tangential method as presented for $P= 0.13$~GPa. For 1.1~GPa, the behavior is rather flat and no transition temperature can be defined in this way. Nevertheless there may be some residual fraction of the ferromagnetic transition. The initial broadness of the anomaly and this special behavior give strong indications that there may be a distribution of Curie temperatures within the sample.

The inset of Fig. \ref{data} gives a zoom on the superconducting transition from resistivity measurements $T_{\rm sc}(\rho)$ for different pressures. Superconductivity has been observed up to the highest measured pressure of $P=2.4$~GPa. The superconducting transition temperature $T_{\rm sc}(\rho)$ is defined by the midpoint of the transition in the resistivity. 

\begin{figure}
\begin{center}
 \includegraphics[width=76mm]{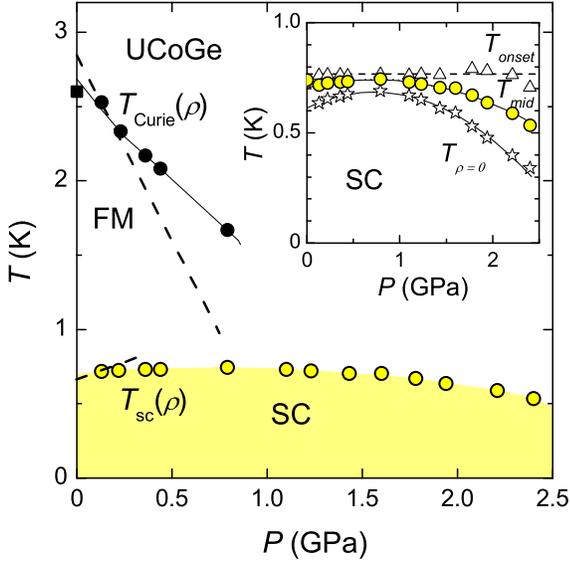}
\end{center}
 \caption{(Color online) Phase diagram $(T, P)$ of UCoGe from resistivity measurements with the ferromagnetic (FM) phase and the superconducting (SC) phase (circles). At zero pressure $T_{\rm Curie}$ is determined by ac susceptibility (full square). The dashed lines indicate the slopes of the transition lines calculated from Ehrenfest relation at ambient pressure \cite{Huy2007}. In the inset the onset-, midpoint- and zero-resistivity temperatures of the superconducting transition are shown. Lines are guides to the eye.}
\label{phasedia}
\end{figure}

\begin{figure}
\begin{center}
\includegraphics[width=60mm]{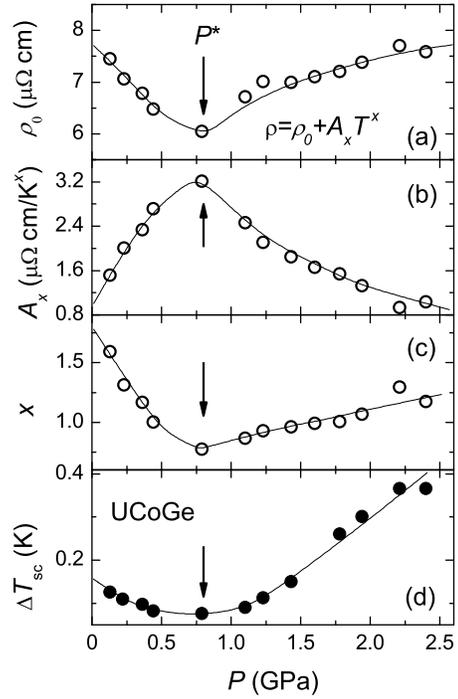}
\end{center}
\caption{(a-c) Pressure dependence of the fit parameters of a $\rho=\rho_0+A_xT^x$ fit in the normal state for $T<$1.7~K in UCoGe. The clear anomaly at $P^{\star}\sim0.8$~GPa reflects the pressure where the broad transition passes through this temperature region. It corresponds to the pressure where the superconducting transition $\Delta T_{\rm sc}$ is the narrowest (d). The lines are guides to the eye.} \label{pressdep}
\end{figure}

The phase diagram obtained from these measurements is shown in Fig. \ref{phasedia}. The dashed lines indicate the slopes of $dT_{\rm Curie}(\alpha)/dP=-2.5$~K/GPa and $dT_{\rm sc}(\alpha)/dP=0.48$~K/GPa as determined from Ehrenfest relation by the specific heat and thermal expansion ($\alpha$) anomaly \cite{Huy2007}. The obtained slopes from this high pressure study are different, $dT_{\rm Curie}(\rho)/dP=-1.4$~K/GPa and $dT_{\rm sc}(\rho)/dP=(0.1\pm0.05)$~K/GPa, but agree in sign. An interpolation of $T_{\rm Curie}(\rho)$ data to ambient pressure gives a $T_{\rm Curie}(\rho)$ of 2.7~K in good agreement with $T_{\rm Curie}$ determined from ac susceptibility (maximum of $\chi_{\rm ac}'$). The inset of Fig. \ref{phasedia} shows the onset ($T_{\rm onset}$), midpoint ($T_{\rm mid}$), and zero resistivity ($T_{\rho=0}$) of the superconducting transition in an enlarged scale as a function of pressure. Under pressure $T_{\rm mid}$ and $T_{\rho=0}$ first increase and then decrease with a maximum of $T_{\rm sc}(\rho)=T_{\rm mid}=0.75$~K at $P=0.8$~GPa. The onset temperature $T_{\rm onset}$, however, depends only weakly on pressure (see Fig. \ref{pressdep}(d)).
At first glance, the UCoGe phase diagram seems different from the ones of UGe$_2$ and URhGe. In contrast to UGe$_2$ and URhGe, the superconducting temperature appears to be almost independent of pressure and insensitive to the transition through $P_{\rm c}$. In first approximation $P_{\rm c}$ may be respectively expected at 2.1~GPa or at 1.2~GPa from a linear extrapolation of the low pressure data of $T_{\rm Curie}(\rho)$  and of $T_{\rm Curie}(\alpha)$(dashed line in Fig. \ref{phasedia}). However, as the transition from the ferromagnetic to the paramagnetic state may be of first order, as generally expected for ferromagnetic quantum phase transitions \cite{Belitz1999, Chubukov2004}. $T_{\rm Curie}$ will not be continuously suppressed to $T=0$ but will be finite at a pressure $P_{\rm c}^{\star}$ above which no ferromagnetism emerges anymore. $P_{\rm c}^{\star}$ is probably below 2.1~GPa. When $T_{\rm sc}$ becomes higher than $T_{\rm Curie}$, ferromagnetism will not survive as the opening of a superconducting gap precludes the establishment of long range magnetic order. The critical pressure $P_{\rm c}^{\star}$ for ferromagnetism  can therefore be estimated from the crossing point between the transition lines from resistivity and thermal expansion with the superconducting transition $T_{\rm sc}(P)$ as $P_{\rm c}^{\star}(\rho)=1.2$~GPa respectively $P_{\rm c}^{\star}(\alpha)=0.8$~GPa. It is interesting to note that $P_{\rm c}^{\star}$ is approximately equal to the characteristic pressure $P^{\star}$ as described later.

Another mark of unusual behavior appears in the analysis of the resistivity data according to the equation $\rho=\rho_0+A_xT^x$ in the normal state for $T<$1.7~K as shown in Fig. \ref{pressdep}. The $T^2$ Fermi-liquid law is only found at low pressure. From the $P$ variation of the residual resistivity $\rho_0$, the $A_x$ coefficient and the derived exponent $x$ of these fits as well as from the apparent estimated broadening $\Delta T_{\sc}$ of the superconducting transition, clearly a characteristic pressure of $P^{\star} \sim 0.8$~GPa emerges. At $P^\star$ the coefficient of the inelastic scattering term $A_x$ is enhanced. A quite unusual result is the quasi-invariance of $x\sim1$ above $P^{\star}$. The difficulty to recover $T^2$ Fermi liquid law on both sides of the first order quantum critical point has now been established in many systems, notably for MnSi \cite{Doiron2003} and ZrZn$_2$ \cite{Takashima2007}. However, $x$ is often very near to 1.5, while here it is more close to one. Of course, we cannot exclude that a Fermi liquid state with a clear $T^2$ temperature dependence of the resistivity appears only at a temperature lower than $T_{\rm sc}$. %
The coincidence under pressure between a maximum of $T_{\rm sc}$ and a minimum in the width of the superconducting transition in the resistivity is a common feature in pressure experiments. However, the fast increase of the broadening of the superconducting transition above $P^{\star}$ may indicate a rapid pressure dependence of the superconducting volume fraction associated with surviving ferromagnetic droplets. This suggests that $P_{\rm c}^{\star}$ may be not so far above $P^{\star}$ and the pressure of the collapse of superconductivity will coincide with $P_{\rm c}^{\star}$ as in UGe$_2$ in high purity crystals.

Due to the long range nature of ferromagnetism, the sample purity may play a key role in the electronic properties close to the ferromagnetic instability. This question becomes critical for a weak first order transition as expected for UCoGe. UGe$_2$ may be a rather ``clean'' as $\Delta m_0=$1~$\mu_{\rm B}$, while in UCoGe $\Delta m_0$ may be near $10^{-2}$~$\mu_{\rm B}$ at the first order transition.

Even the previous reports point out the heterogeneity of the phase transition at $T_{\rm sc}$ but also at $T_{\rm Curie}$. For example in the first publication about UCoGe superconductivity, the resistivity onset of superconductivity at $T_{\rm sc}=0.8$~K is far above the maximum of the superconducting specific heat anomaly at $T_{\rm sc}\sim 0.45$~K \cite{Huy2007}. The difficulty to achieve a homogeneous state may also be indicated in NMR and NQR results (weak fraction of superconducting volume, no simple exponential behavior of nuclear magnetization recovery) \cite{Ohta2008}.

If there is heterogeneity in ferromagnetism, the resistivity probe is not adapted to test the boundary of bulk superconductivity because a remaining fraction of ferromagnetism may lead to partial superconductivity in the sample and hence an apparent weak pressure dependence of $T_{\rm sc}$ detected by resistivity. As recently demonstrated for the uranium-based heavy fermion superconductor URu$_2$Si$_2$, the superconducting phase boundary obtained by resistivity measurements can be very different from that measured by specific heat \cite{Hassinger2008}. Furthermore in URu$_2$Si$_2$, the collapse of bulk superconductivity is at $P_{\rm x}= 0.5$~GPa, when the hidden order phase switches to the antiferromagnetic ground state, while resistivity data indicate the suppression of superconductivity between 1.3~GPa and 2~GPa.

Experimentally the surprising observation was the difficulty to modify the superconducting state under pressure. However we believe that this unexpected behavior is linked to the simultaneous high sensitivity of ferromagnetic and superconducting anomalies to imperfections. Here it is shown that even for a $RRR$ near 30, large smearings occur in ferromagnetic and superconducting transitions. In heavy fermion systems with an antiferromagnetic instability, the smearing appears mainly for the superconducting transition. Furthermore, it is interesting to note that in contrast to antiferromagnetically ordered systems like CeRhIn$_5$ \cite{Knebel2006} or CeIrSi$_3$ \cite{Tateiwa2007}, the superconducting transition in UCoGe is sharper in the magnetically ordered regime than in the paramagnetic state.

There are strong indications for a critical behavior at $P^{\star}\sim 0.8$~GPa being almost the critical pressure of $P_{\rm c}^{\star}$ predicted from ambient pressure thermal expansion measurements. The apparent strong deviations from Fermi liquid behavior above $P^{\star}$ is a remarkable fact which may be due to a surviving ferromagnetic cluster. 

In conclusion, we present a high pressure phase diagram of the ferromagnetic superconductor UCoGe determined from resistivity measurements on a polycrystalline sample.  The ferromagnetic transition temperature is monotonously suppressed with pressure and could be followed up to $P^{\star}\sim 0.8$~GPa. The pressure dependence of $T_{\rm sc}$ has a smooth maximum at $P^{\star}$ and superconductivity could be observed up to the highest pressure of 2.4~GPa. However, the broadening of the superconducting transition above $P^{\star}$ may indicate an inhomogeneous state. Obviously the next target will be to perform measurements on high quality single crystals ($RRR>50$).

\section*{Acknowledgment}
We thank J.-P. Brison and A. Miyake for stimulating discussions and D. Braithwaite for experimental support. This project was financially supported by French ANR projects ECCE and CORMAT.


\end{document}